\documentclass[11pt]{article}

\usepackage[utf8]{inputenc}
\usepackage[T1]{fontenc}
\usepackage{amsmath,amssymb,amsthm,bbm}
\usepackage{graphicx,transparent,eso-pic}
\usepackage[document]{ragged2e}
\usepackage{natbib}
\usepackage{multicol}
\usepackage{url}
\usepackage[margin=2cm]{geometry}

\newcommand{\noun}[1]{\textsc{#1}}

\usepackage{natbib}

\begin{document}

\title{Hierarchy and co-evolution processes in urban systems}
\author{\noun{Juste Raimbault}$^{1,2,3}$\\
$^1$ CASA, UCL\\
$^2$ UPS CNRS 3611 ISC-PIF\\
$^3$ UMR CNRS 8504 G{\'e}ographie-cit{\'e}s\\
\texttt{juste.raimbault@polytechnique.edu}
}
\date{}

\maketitle

\justify

\begin{abstract}
The concept of hierarchy in complex systems is tightly linked to co-evolutionary processes. We propose here to explore it in the case of the co-evolution between transportation networks and territories. More precisely, we extend a co-evolution model for systems of cities and infrastructure networks, and systematically study its behavior following specific hierarchy indicators we introduce. We show that population hierarchy and network hierarchy are tightly linked, but that a broad range of regimes can exist. Model exploration furthermore yields non-trivial stylized facts which can be taken into account for territorial planning on such long time scales with co-evolutionary processes.
\end{abstract}


\section{Introduction}

\subsection{Complexity and hierarchy}

Complex systems with emergent properties produced by self-organization processes are also most of the time exhibiting some kind of hierarchical structure. Although the term of hierarchy has several different definitions and uses in very different disciplines, ranging from political science \citep{crumley1987dialectical} to physics \citep{10.1371/journal.pone.0033799}, it seems to be intrinsically linked with complexity. \cite{lane2006hierarchy} classifies four frequent uses of the term hierarchy, namely (i) order hierarchy corresponding to the existence of an order relation for a set of elements, (ii) inclusion hierarchy which is a recursive inclusion of elements within each other, (iii) control hierarchy which is the ``common sense'' use of the term as ranked entities controlling other entities with lower rank, and (iv) level hierarchy which captures the multi-scale nature of complex systems as ontologically distinct levels (or scales). For the particular study of social systems, he concludes that hierarchical levels may be entangled, that upward and downward causations are both essential, and that at least three levels (micro, meso, macro) are generally needed to capture the complexity of such systems. In a more philosophical account of complexity, \cite{morin1980methode} constructs a hierarchical method of interdisciplinary knowledge, insists on the tension between dependancy and interdependency or between opening and closing (rejoining ideas from \cite{holland2012signals}), and develops an implicit hierarchy of social systems when hypothesizing the emergence of third-type societies (swarm intelligence between humans).

Different types of complexity may be related to different types of hierarchy as \cite{raimbault:halshs-02089520} proposes, and hierarchy would indeed be endogenous to theories of complexity. \cite{allen2017multiscale} develop a multiscale information theory in which the information profile across scales, or hierarchical levels, allows quantifying the complexity of a system. The complex adaptive system theory of \cite{holland2012signals} considers complex systems as systems of boundaries that filter signals, implying an inclusion and scale hierarchy between boundaries. Theories of scaling as the one synthesized by \cite{west2017scale} rely on the quantification of hierarchy in certain dimensions of systems, captured by exponents of scaling laws. Hierarchy may be endogenous to complexity, or to knowledge of the complex itself, since for example \cite{fanelli2013bibliometric} provides empirical evidence of a ``hierarchy of sciences'', in the sense of possibility to reach theoretical and methodological consensus. This corresponds in some sense to the ``ontological complexity'' of \cite{pumain2003approche}, which relies on the number of viewpoints needed to grasp a system, or the number of perspectives in an applied perspectivism framework \citep{raimbault2020relating}. Wether linked to systems themselves or to models and theories of them, hierarchy appears to be tightly linked to complexity.

\subsection{Territorial systems and hierarchy}

Urban systems, and more generally territorial systems, are particularly linked to hierarchy \citep{pumain2006hierarchy}: they indeed encompass all the meanings aforementioned (order hierarchy between settlement sizes for example, inclusion hierarchy between territorial boundaries, control hierarchy through governance structure, and more importantly level hierarchy through their multi-scalar nature). \cite{batty2006hierarchy} shows that hierarchies are inherent to urban systems, as fat tail distribution of settlement size are already produced by simple models of urban growth, and suggests also that urban design processes imply underlying overlapping hierarchies. \cite{pumain2006alternative} links hierarchical selection and hierarchical diffusion of innovation across cities to the long-term dynamics of urban systems. \cite{pumain:halshs-02303136} recalls that interactions in systems of cities are tightly linked to the emergence of urban hierarchies. Generally, scaling laws in urban systems can be considered as systematic manifestations of a hierarchical structure \citep{pumain2004scaling}, which is more complex than a simple order hierarchy, since scaling patterns vary with the definition of cities \citep{cottineau2017diverse}

Hierarchical properties can be observed on several dimensions of urban systems. For example, transportation systems are hierarchical in their structure \citep{yerra2005emergence} but also patterns of use such as transportation flows \citep{jiang2009street}. Urban hierarchies are tightly related to hierarchies of their transportation links \citep{bigotte2010integrated}, and different modes of transportation networks are concerned including the air network \citep{dang2012hierarchy}. The global distribution of multinational firms also exhibits strong hierarchical patterns \citep{godfrey1999ranking}. Governance structures are organized following both an inclusion hierarchy for administrative areas \citep{li2015administrative} but also level hierarchies for example for economic processes \citep{liao2017opening}. Territorial systems are therefore intrinsically hierarchical in their multiple dimensions, what is tightly linked to their different complexities \citep{2019arXiv190109869R}.

\subsection{Co-evolution and hierarchy}

Hierarchy in complex systems is furthermore intrinsically linked to the concept of co-evolution. Following \cite{lane2006hierarchy}, the approach to complex adaptive systems proposed by \cite{holland2012signals} integrates levels and nested hierarchies, since it considers complex systems as ensembles of boundaries that filter signals. \cite{holland2012signals} formalizes complex adaptive systems as these structures of boundaries which form co-evolution niches for the elements and subsystems within a given boundary. The concept is slightly different from the concept of ecological niche which more generally designates a region in a parameter space quantifying the environment in which a species can live. In ecology, \cite{pires2011food} show that the emergence of mutualistic species networks imply some feeding hierarchy.

In the context of economic and geographical processes, \cite{volberda2003co} distinguish for the co-evolution of firms between a genealogical hierarchy (evolutionary processes in the biological sense) and an ecological hierarchy (co-evolutionary economic processes). \cite{liu2013exploring} suggest that air networks co-evolve with firm networks and that their hierarchies are related therethrough. \cite{raimbault2019modeling} introduces a co-evolution to study interactions between transportation networks and territories, which from an urban system viewpoint in the sense of \cite{pumain2006evolutionary} relates to urban hierarchies. \cite{levinson2007co} confirm a correspondance between urban and network hierarchies in a co-evolution model. Within the SimpopNet model for the co-evolution of cities and networks \citep{schmitt2014modelisation}, discrete hierarchical level of network links corresponding to successively improved transportation technologies are a core component of simulation rules. \cite{raimbault2020unveiling} furthermore showed that the level of initial urban hierarchy in terms of rank-size slope had significant impacts on model outcomes. Studying hierarchies in the context of co-evolution transportation networks and territories is thus a relevant entry to the underlying concepts, including complexity, hierarchy, co-evolution and territorial systems.

\subsection{Proposed approach}

\cite{pumain2006introduction} recalls in the context of social systems some remaining open methodological questions: how are hierarchies produced? How do hierarchies evolve? What discriminates between continuous and discrete hierarchical organisations? Our contribution brings new elements of answer to the first two questions above, in the particular case of co-evolution of transportation networks and territories. It situates at the intersection of the three previously given contexts, namely hierarchy in complex systems and more particularly territorial systems, seen through the prism of co-evolutive processes.

More precisely, we propose to systematically explore a macroscopic co-evolution model for cities and networks, and study its properties regarding both hierarchies of each components, in terms of final hierarchy produced but also in terms of the relations between these hierarchies. Establishing links between microscopic processes and emergent hierarchical patterns through model exploration informs on possible drivers of these macroscopic patterns. Our contribution relies on three aspects: (i) we introduce a comprehensive set of indicators tailored to the study of hierarchy in territorial systems; (ii) we systematically explore the version with a physical network of the co-evolution model introduced by \cite{raimbault2018modeling} which only studied extensively the virtual network; and (iii) we apply a novelty search algorithm to establish the feasible space of hierarchy patterns which can be produced by the model.

The rest of this chapter is organized as follows. We first describe the model used and introduce a novel set of indicators to quantify hierarchy in territorial systems. We then describe the results of a grid exploration of the co-evolution model using these indicators, both with the physical and virtual networks, and establish the feasible space of model outputs. We finally discuss the implications of these results for hierarchy within co-evolutionary processes.

\section{Co-evolution model}

\subsection{Context}

The issue of interactions between transportation networks and territories remains an open question for which different approaches have been proposed \cite{offner1993effets,espacegeo2014effets}. \cite{raimbault2018caracterisation} has explored a co-evolution approach, in the sense that both dynamics have circular causal relationships. More precisely, \cite{raimbault2019modeling} introduces a definition of co-evolution in that particular context, based on the aforementioned co-evolution niches \citep{holland2012signals}, for which an empirical characterization method based on lagged correlations is developed \citep{raimbault2017identification}. As its application on empirical data yield various or inconclusive results, the use of simulation models is a medium to indirectly link microscopic processes with a potentially emergent co-evolution, both at the mesoscopic scale \citep{raimbault2019urban} and at the macroscopic scale \citep{raimbault2018modeling}. This latest model is the one used in this study.

\subsection{Model description}

The co-evolution model for cities and transportation networks at the macroscopic scale extends the spatial interaction model introduced by \cite{raimbault2018indirect} by adding dynamical speeds to network links. A system of cities is represented as cities as agents and network links between these. Interaction flows are determined with a spatial interaction model, and they determine city growth rates, while network links evolve according to the flow traversing them. See \cite{raimbault2018modeling} for a full mathematical description of the model. We describe below the specification and parameters used here.

More precisely, a time step of the simulation model consists in the following steps:
\begin{enumerate}
	\item cities evolve their populations following gravity interaction flows of unit weight $w_G$, as a scaling function of population with exponent $\gamma_G$, and with a distance decay parameter $d_G$; cities do not have endogenous growth in our setting (Gibrat model) as we focus our study on interactions;
	\item flows are assigned to network links, either (i) to the direct link between the two cities in the case of the virtual network, or (ii) through a shortest path assignment algorithm (betweenness centrality) in the case of a physical network;
	\item links evolve their speed with a thresholded self-reinforcement function of flows, with maximal time travel decrease $g_M$; a threshold for flows above which (resp. below which) speed increase (resp. decrease) determined by a flow quantile parameter $\phi^{(q)}_0$; and as a scaling relation to relative flows with exponent $\gamma_N$.
\end{enumerate}

\begin{figure}
	\includegraphics[width=0.95\linewidth]{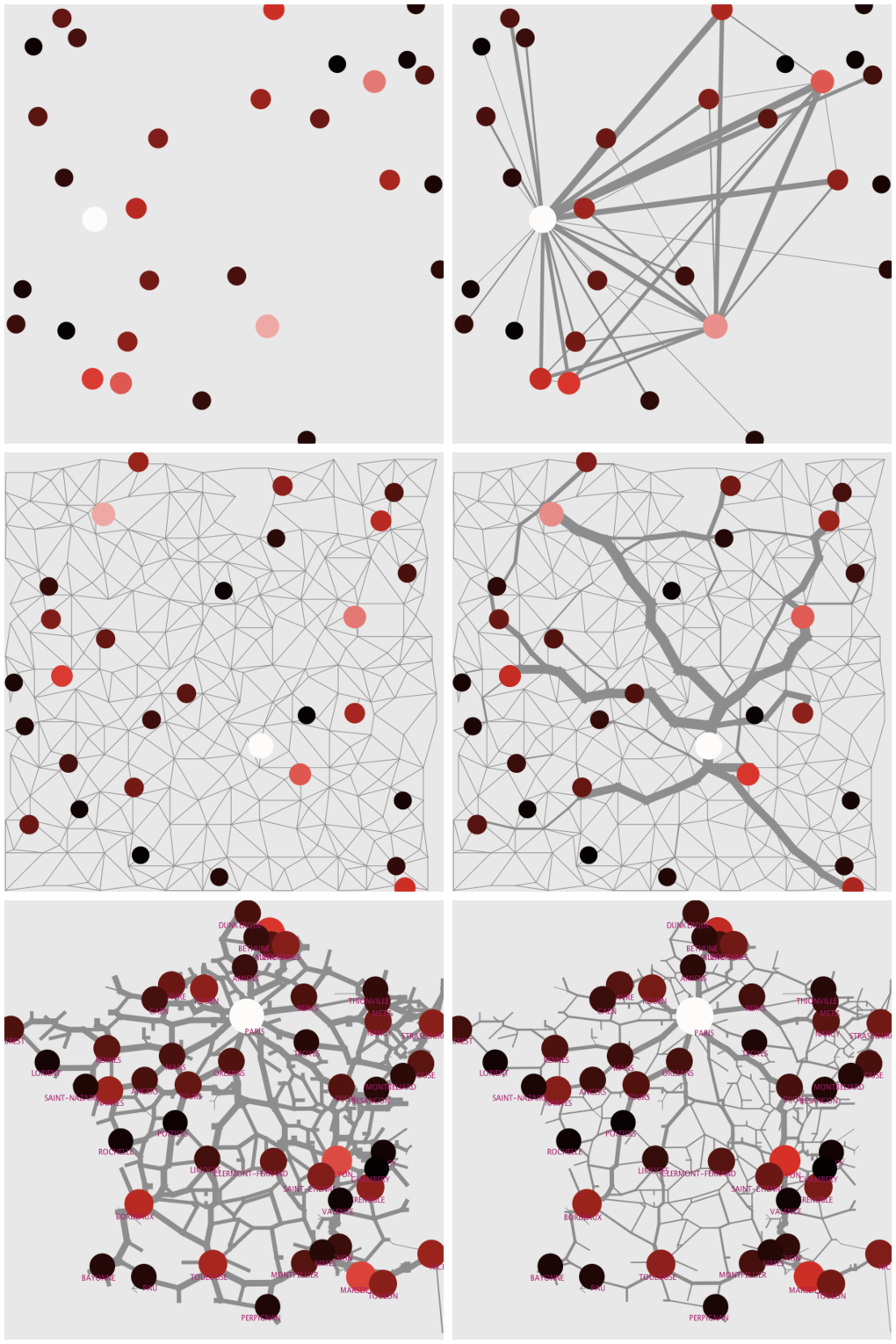}
	\caption{\footnotesize\textbf{Examples of different setups for the co-evolution model.} \textit{(Top row)} Synthetic system of cities with virtual network, initial configuration (left) and after $t_f=30$ time steps (right), with parameters $\alpha_S = 1$, $\phi^{(q)}_0 = 0.9$, $g_M = 0.01$, $\gamma_N=2$, $w_G=4.7e-3$, $d_G=248$, $\gamma_G=0.9$; \textit{(Middle row)} Synthetic system of cities with physical network, initial configuration (left) and after $t_f=30$ time steps (right), with parameters $\phi^{(q)}_0 = 0.7$, $g_M = 0.05$ and the same other parameters than the first configuration; \textit{(Bottom row)} French system of cities simulated between 1975 (left) and 1999 (right) with three time steps, with parameters $\phi^{(q)}_0 = 0.8$, $g_M = 0.2$, $\gamma_N=4$ and same others. City color and size give the population and link thickness the speed (rescaled at each time step).\label{fig:examples}}
\end{figure}

The model can be initialized with real data or by generating a synthetic initial configuration which has its own parameters \citep{raimbault2019space}. In our case, $N = 30$ cities are randomly distributed in an uniform space of width $W=200km$, and population follow a rank-size law with parameter $\alpha_S$. For the virtual network case, all pari links are initialized with pace one, while for the physical network case a perturbed grid network is used as described in \cite{raimbault2018modeling}. We show in Fig.~\ref{fig:examples} model runs for synthetic virtual and physical networks, and for the French system of cities with railway data. We visually observe that the number of important links is smaller in the physical case as it could be expected as infrastructure is shared by neighboring flows. For the real system, the most important links emerging correspond roughly to the actual existing high-speed lines.

In our exploration settings, the model has thus six parameters (for which we give practical boundaries in experiments): the initial population hierarchy $\alpha_S \in \left[0.1; 2.0\right]$, the gravity interaction weight $w_G \in \left[1e-4; 1e-2 \right]$, the gravity interaction hierarchy $\gamma_G \in \left[0.0 ; 5.0 \right]$, the gravity decay $d_G \in \left[1.0; 500.0 \right]$, the network maximal speed growth $g_M \in \left[0.0; 0.05 \right]$, the network growth hierarchy $\gamma_N \in \left[0.0; 5.0\right]$, and the network threshold quantile $\phi_0^{(q)} \in \left[0;1\right]$.

\subsection{Quantifying hierarchy in systems of cities}

Indicators to understand macroscopic trajectories in simulated systems of cities have been introduced by \cite{raimbault2020unveiling}. They include some related to hierarchy but are not specifically focused on this aspect. We propose now to give a broad set of indicators to capture different dimensions of hierarchy.

\subsubsection{Static quantification of hierarchy}

The most straightforward way to quantify hierarchy is to use Zipf rank-size law in the case of population, or more generally scaling laws for other dimensions of the urban system. Let $Y_i$ the variable for which the hierarchy is estimated. Assuming $i$ is ordered in decreasing order, the Ordinary Least Square estimation of $\log \left(Y_i\right) \sim \log \left( i\right)$ gives an estimation of the rank-size slope $\alpha \left[Y\right]$ which is a proxy of hierarchy. Additional indicators to explain more accurately the size distribution include for example the primacy index. We take a generic approach to this issue of more degree of freedoms to capture the distribution and use a piecewise linear regression, implementing the algorithm of \cite{muggeo2003estimating}. Given the distribution observed empirically and the ones generated by simulation models, going beyond one breakpoint does not bring significant improvement. We consider thus the estimated slopes and breakpoint as refined indicators of the hierarchy, given as $\alpha_1 \left[Y\right]$,  $\alpha_2 \left[Y\right]$ and  $\Psi \left[Y\right]$. Finally, to quantify interactions between two aspects, a correlation between two hierarchies informs they correspond in terms of ranks, and is computed with $r_s\left[X_i,Y_i\right]$ for two variables $X_i,Y_i$ with $r_s$ an estimator of Spearman rank correlation.

\subsubsection{Dynamical indicators}

The rank correlation between initial and final distribution of a variable will measure how much an ordering hierarchy was modified, which is different from the variation of hierarchy given the variations of previous indicators such as the rank-size slope. Dynamical indicators for hierarchy regimes can furthermore be defined in several ways: dynamics of the rank correlation between two variables, time-series properties of rank-size trajectories, lagged rank correlations. Studying these extensively is out of the scope of this chapter, and we will consider differences between initial and final hierarchies to capture dynamics.

\subsubsection{Spatialized indicators}

Finally, some spatial extension of hierarchy indicators can be introduced. A spatial non-stationary version of a scaling law would write $Y_i (\vec{x}) \sim \left(\frac{X_i(\vec{x})}{X_0 (\vec{x})}\right)^{\alpha (\vec{x})}$, where $\vec{x}$ is the spatial position and assuming that samples can be defined at each point in space. In practice, a discrete version could be more relevant, for which $\vec{x}_k$ center point are defined, samples consist of points within Thiessen polygons of centers and the exponents are estimated for each center $\alpha (\vec{x}_k)$. Some heuristics should be developed to estimate such a discrete non-parametric scaling law, and also remains out of our scope here.

\section{Results}

\subsection{Implementation}

The model is implemented in NetLogo \cite{tisue2004netlogo}, which is a good compromise between performance and interactivity, the former being necessary with a model with such a spatialized network. The model is explored using the OpenMOLE model exploration software \cite{reuillon2013openmole} to use integrated design of experiments (DOE) and exploration methods, but also the seamless access its provide to high performance computing infrastructures. Source code for model, exploration scripts, result analysis, and results are available on a git repository at \texttt{https://github.com/JusteRaimbault/CoevolutionNwTerritories}. Large dataset for simulation results are available on the dataverse at \texttt{https://doi.org/10.7910/DVN/6GUKOX}.

\subsection{Hierarchy patterns}


We now turn to a first basic exploration of the model, using a grid exploration of the parameter space. A first broad grid varying all parameters with 3 steps each and 20 model repetitions, for both virtual and physical models,  allows identifying the dimensions along which no significant variation or qualitative variation in model behavior occur. We then fix a more targeted exploration by taking $g_M = 0.05$ and $\gamma_N = 1$ and varying $\alpha_S \in \{0.5, 1.0, 1.5 \}$, $\phi_0^{(q)} \in \{0.1, 0.5, 0.9 \}$, $\gamma_G \in \left[0.5;1.5\right]$ with a step of 0.2, and $d_G \in \left[10; 210 \right]$ with a step of 50. We consider the static indicators of hierarchy and their variation between initial and final time, applied on city populations $P$ and city closeness centrality $C$.

\begin{figure}
\includegraphics[width=\linewidth]{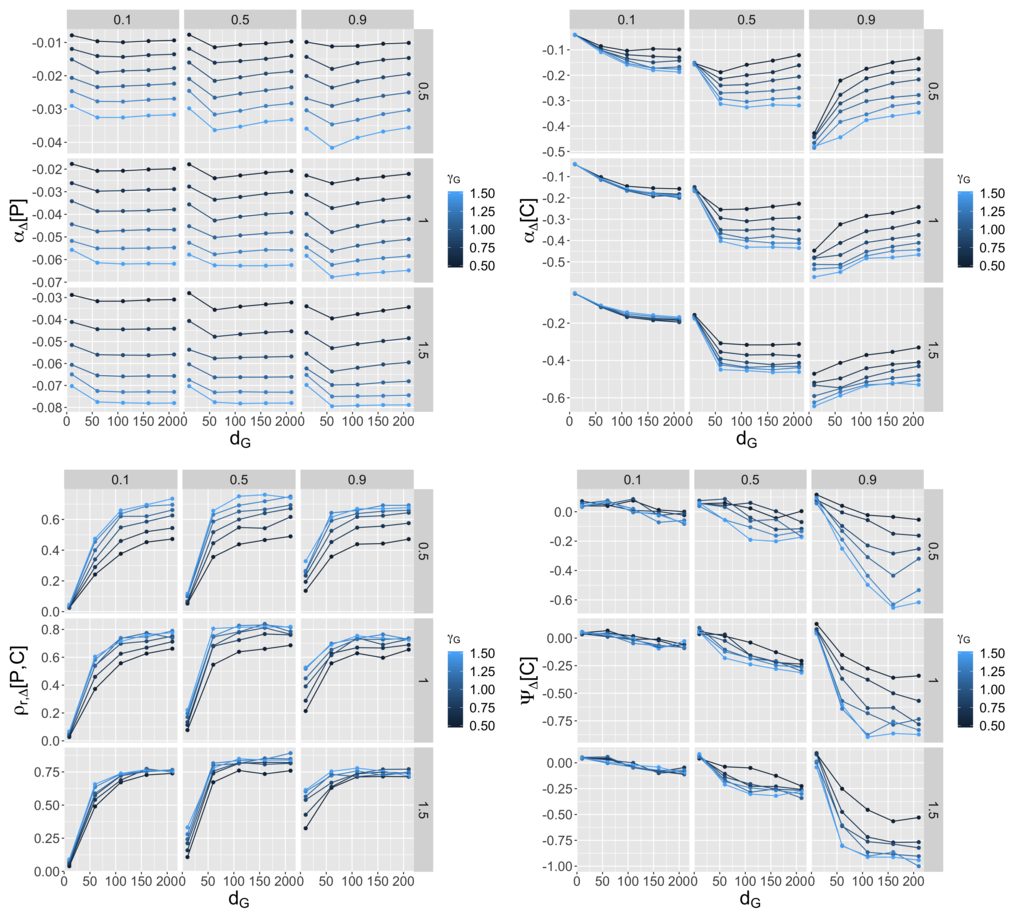}
\caption{\textbf{Patterns of hierarchy in the model with a virtual network.} Each indicator is shown as a function of $d_G$ for varying $\gamma_G$ (color), varying $\phi_0^{(q)}$ (columns) and varying $\alpha_S$ (rows). \textit{(Top Left)} Difference in the rank-size exponent for populations between final time and initial time; \textit{(Top Right)} Difference in the rank-size exponent for centralities; \textit{(Bottom Left)} Difference in rank correlation between population and centralities; \textit{(Bottom Right)} Difference in breakpoint of the hierarchy of centralities. \label{fig:gridexplo-virtual}}
\end{figure}

The variation of some indicators exhibiting an interesting behavior are shown for the model with the virtual network in Fig.~\ref{fig:gridexplo-virtual}. The evolution of city population hierarchy, captured by $\alpha_{\Delta}\left[P\right] = \alpha \left[P\right](t_f) - \alpha_S$ (top left panel of Fig.~\ref{fig:gridexplo-virtual}), exhibits a low qualitative sensitivity to initial hierarchy $\alpha_S$ (rows), but subplots are translated and a significant quantitative sensitivity is observed: in other words, more hierarchical systems produce more hierarchy, what can be expected in such self-reinforcing processes. Always negative values mean that hierarchy always increases. As a function of gravity decay $d_G$, a systematic absolute decrease is observed for the lowest values: very local interaction mitigate the increase in hierarchy. The gravity interaction $\gamma_G$ has a monotonous and expected effect, systematically increasing the hierarchy. Finally, an effect of the co-evolution with network distances which yields non-monotonous effects is worth noticing: when network threshold $\phi_0^{(q)}$ increases, a minimum of $\alpha_{\Delta}\left[P\right]$ is observed for high $\gamma_G$ values and low initial hierarchy. In that context, an intermediate range of spatial interaction will yield more hierarchical systems. As only few links increase their speed with this value of network threshold, it means that long range interactions are no longer amplified by the network. Thus, the evolution of city hierarchies depends on several parameters, in a non-monotonous way when interacting with network processes.

Regarding the evolution of network hierarchies $\alpha_{\Delta}\left[C\right]$ (top right panel of Fig.~\ref{fig:gridexplo-virtual}), the most significant effect is the one of network threshold $\phi_0^{(q)}$, which witnesses an inversion of the sense of variation as a function of distance decay $d_G$ when network threshold increases. When all links are allowed to grow their speed, longer span interaction will lead to more hierarchical networks: indeed, the probability for two large cities to interact is then higher, and their flow will be favored in terms of network growth. But when only a few proportion of links improve their travel time while most of them decay, then higher network hierarchies are produced by the most local interactions. In a setting of a scarcity of network investments, taking into account long range interaction gives thus a more balanced network than a local approach, which is kind of counter-intuitive.

The behavior of the rank correlation between population and centrality $\rho_r \left[P, C \right]$ (bottom left panel of Fig.~\ref{fig:gridexplo-virtual}) informs on the co-evolution processes between the territory and the transportation network. Empirically, the better connectivity of larger cities has been suggested by \cite{bretagnolle2003vitesse} as a signature of co-evolution processes. Our results confirm that indeed such co-evolution processes produce a correspondance between city and network hierarchy, as high values of correlations are attained for interaction spans over 100km. The interaction distance furthermore systematically increase the correlation, and local interaction yield a close to zero correlation except for highly initially hierarchic systems with a high network threshold (in which case some very large cities will still construct a local interaction system). The correlation is maximal at an intermediate value of $\phi_0^{(q)}$, which means that the link selection process plays a role in the synchronization between the two hierarchies, and that the co-evolution process captures more than just a self-reinforcement.

Finally, as we introduced segmented regression as a finer characterization of hierarchical patterns in a system of cities, we observe an interesting behavior for the variation of the breakpoint for centralities $\Psi_{\Delta}\left[C\right]$. The breakpoints always shifts in time to lower values, meaning that the distribution becomes more unequal in time regarding the most dominating links (in the sense that less links are included in the head of the hierarchy). The shift is stronger when interaction distance is larger and network threshold is larger, meaning that favoring less long range links will break the hierarchy in a more uneven way.

\begin{figure}
\includegraphics[width=\linewidth]{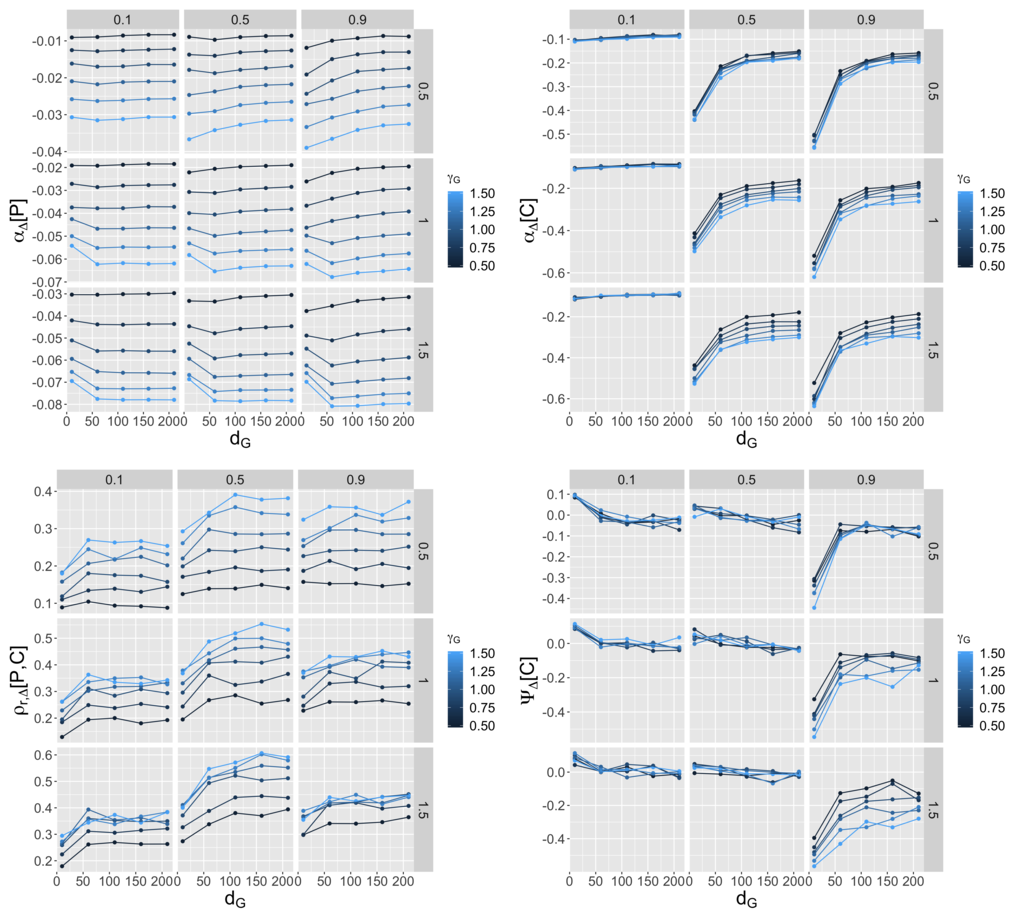}
\caption{\textbf{Patterns of hierarchy in the model with a physical network.} With the same design of experiment than in Fig.~\ref{fig:gridexplo-virtual}, each indicator is shown as a function of $d_G$ for varying $\gamma_G$ (color), varying $\phi_0^{(q)}$ (columns) and varying $\alpha_S$ (rows). \textit{(Top Left)} Difference in the rank-size exponent for populations between final time and initial time; \textit{(Top Right)} Difference in the rank-size exponent for centralities; \textit{(Bottom Left)} Difference in rank correlation between population and centralities; \textit{(Bottom Right)} Difference in breakpoint of the hierarchy of centralities.\label{fig:gridexplo-physical}}
\end{figure}

Our second experiment to study patterns of hierarchy is the same exact Design of Experiment than the previous one, but with the physical network. We show in Fig.~\ref{fig:gridexplo-physical} the same indicators for the same parameter space. The discrepancy between the two behaviors is particularly relevant from a thematic point of view, as it reveals the role of spatializing and assigning network flows, even in such as case where no congestion is included. Some patterns are similar, but some important differences can be observed. Globally, the behavior of population hierarchy, rank correlation, and centrality hierarchy breakpoint, are qualitatively similar. The minimum which existed for population hierarchies at short range interactions mostly disappears (although still slightly present for $\gamma_G = 1.5$, $\phi_0^{(q)} = 0.9$ and $\alpha_S = 1$): spatializing the network removes some output complexity in that case. Rank correlations (bottom panel of Fig.~\ref{fig:gridexplo-physical}), i.e. the correspondance between population and centrality hierarchies, is still growing as a function of $d_G$ and exhibits a maximum at the intermediate value $\phi_0^{(q)}$. However, the effect of interaction hierarchy $\gamma_G$ is much more impactful in this case: more uniform interactions (low $\gamma_G$) lead to a much smaller correlation. This means that the approximation of using a virtual network accurately captures the hierarchy correspondance in physical networks for flows with a superlinear scaling exponent: depending on the type of activities generating flows, spatial structure of the network is more or less important. 

The hierarchy of centralities also behave quite differently when switching to a physical network (top right panel of Fig.~\ref{fig:gridexplo-physical}). It is in that case roughly insensitive to any parameter when all links are growing ($\phi_0^{(q)} = 0.1$), and always grows as a function of distance decay $d_G$: longer range interactions diffuse through most network links and yield less inequality between their speeds. Increasing the hierarchy of interactions still increases the hierarchy of centralities but the effect is less strong. In a nutshell, constraining spatially the link and making them share flows through the assignment procedure restricts the degrees of freedom their speed dynamics have.

\subsection{Hierarchy regimes}



\begin{figure}
	\includegraphics[width=\linewidth]{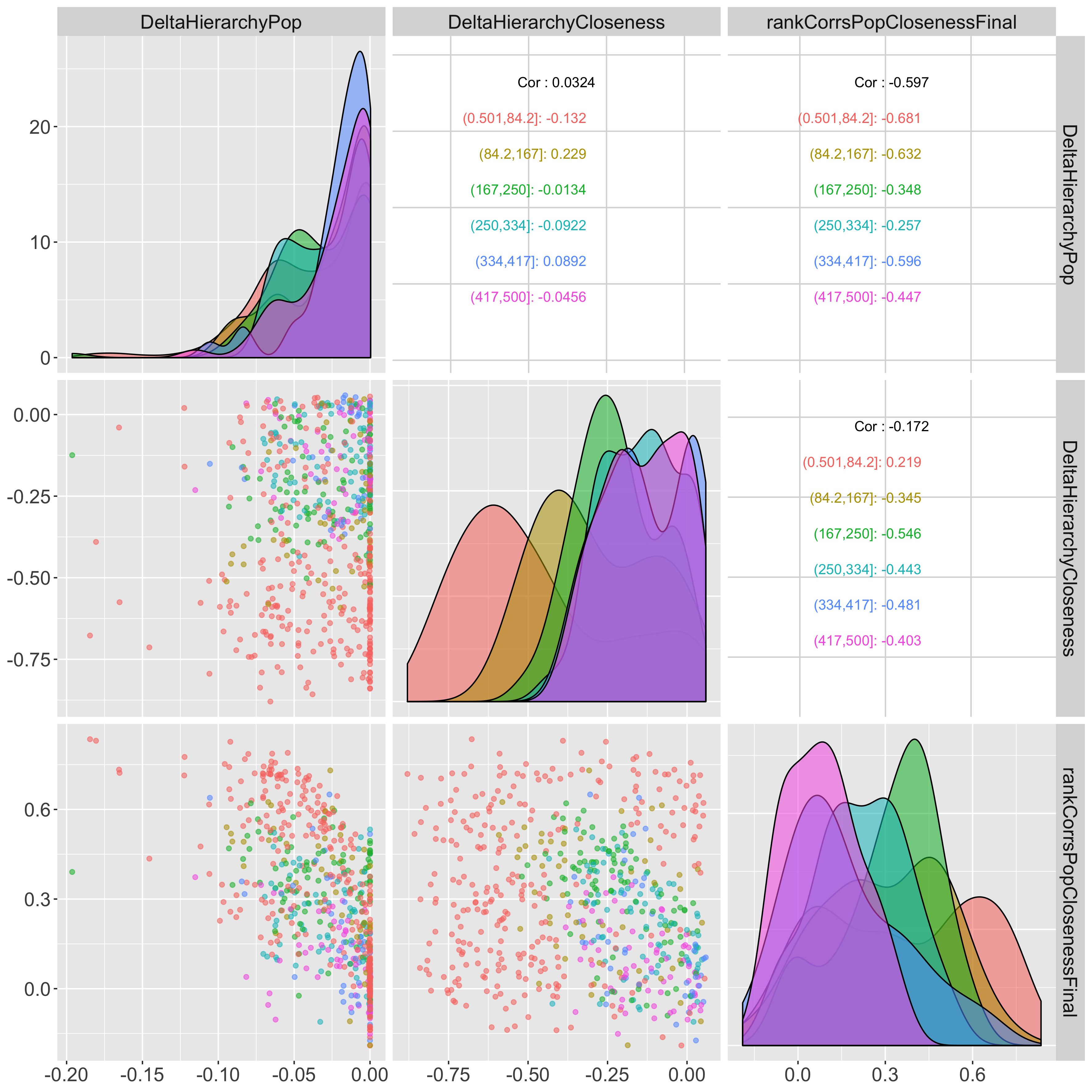}
	\caption{\textbf{Feasible space of hierarchy regimes obtained with the PSE algorithm.} Scatter plots of the three objective dimensions. Color level gives the value of $d_G$, and distributions and correlation between indicators are also stratified following $d_G$. Patterns were filtered to have at least ten stochastic repetitions.\label{fig:pse}}
\end{figure}

After having inspected the link between parameters and hierarchical patterns emerging through a basic grid exploration, we turn now to a specific experiment aimed at establishing the feasible hierarchy regimes that the model can produce. Indeed in such complex simulation models, simple DOE may only capture a part of potential behavior, and miss strong non-linearities. Therefore, the Pattern Space Exploration (PSE) algorithm was introduced by \cite{cherel2015beyond} as an heuristic to approximate the feasible space of a model output, based on a novelty search algorithm \citep{lehman2008exploiting}. We apply here this algorithm with the following three dimensional pattern space: evolution of population hierarchy $\alpha_{\Delta}\left[P\right]$, evolution of centrality hierarchy $\alpha_{\Delta}\left[C\right]$, and final rank correlation between population and centrality hierarchies $\rho_r\left[P,C\right]$. For the first two, looking at dynamics is important to control for the artificial initial level of hierarchy $\alpha_S$ for population, while initial centrality hierarchy is solely linked to geometry and exhibits a narrow peak distribution of average $-0.2$ (similar pattern for virtual and physical, the physical distribution being a bit wider). These three dimension capture not only which hierarchies are produced along the two aspects included in the model, but also which relation they have in term of rank correlation.

We run the PSE algorithm using OpenMOLE and distribute the computation on a grid with an island scheme. The grid for patterns, set from previous exploration results, is taken as $\alpha_{\Delta}\left[P\right] \in \left[-0.2;0.2\right]$ with step $0.02$, $\alpha_{\Delta}\left[C\right] \in \left[-0.2;0.2\right]$ with step $0.1$, and $\rho_r \left[P,C\right] \in \left[-1.0,1.0\right]$ with step $0.1$. Variable parameters are aforementioned model parameters, with the addition of $g_M \in \left[0.0;0.05\right]$. The algorithm was run on 500 parallel islands (termination time 10 minutes), for $30,000$ generations (reasonable convergence in terms of number of patterns discovered.

We show in Fig.~\ref{fig:pse} the scatterplot of the obtained feasible space, conditionally to having at least 10 stochastic repetitions (robust patterns). We find that closeness hierarchy dynamics have a much wider range of possible values than population hierarchy dynamics, confirming what was obtained with the grid experiment. Furthermore, possible correlations have also a large span from -0.19 to 0.84, which means that the model can combine the production of a broad set of hierarchies for population and network, but also of their correlations. These correlations take mostly positive values as expected (mutual reinforcement of hierarchies), but are sometimes uncorrelated and can even be negative: in such a setting the lowest cities of the urban hierarchy have the highest centralities. These correspond to a very low initial hierarchy ($<\alpha_S>=0.18$ where the average is taken for points with a negative correlation), a high network reinforcement exponent ($<\gamma_N>=3.2$), a low interaction hierarchy ($<\gamma_G>=0.88$), and long range interactions ($<d_G>=228$). This can be interpreted as diffuse and uniform interaction in a low-hierarchical system which are mostly dominated by network processes. We can also observe in Fig.~\ref{fig:pse} for the $(\alpha_{\Delta}\left[C\right],\rho_r\left[P,C\right])$ point cloud, that around 75\% of the surface covered is by short range interactions, and correspond to extreme values: normal range interactions produce a restricted output space. Finally, it is interesting to note the upper and lower boundaries of the $(\alpha_{\Delta}\left[P\right],\rho_r\left[P,C\right])$ point cloud: population hierarchy increase fixes both kind of linear upper and lower bounds on correlations: high absolute increase of hierarchy imply high correlations, while correlations can not be too high for small variations of population hierarchy. Altogether, this experiment show a high diversity of hierarchy regimes that the model can produce.

\begin{table}
\caption{Linear regression analysis of model behavior based on PSE patterns. Each model is estimated with a Weighted Least Square, with weights being the number of stochastic samples. Significance levels: (***) $p \simeq 0$; (*) $p < 0.01$; () $p > 0.1$.\label{tab:regpse}}
\centering
	\begin{tabular}{@{\extracolsep{5pt}}|l|cc|cc|cc|}
\hline
Model & $\alpha_{\Delta}\left[P\right]$ & &  $\alpha_{\Delta}\left[C\right]$ & & $\rho_r\left[P,C\right]$ & \\ 
\hline 
  Constant & $1.04\cdot 10^{-2}$ & *** & $0.15$ & *** & $-0.27$ & *** \\
  $\alpha_S$ & $-7.2\cdot 10^{-3}$ & *** & $-6.9\cdot 10^{-3}$ & & $-1.4\cdot 10^{-2}$ & \\
  $\phi_0^{(q)}$ & $-8.6\cdot 10^{-4}$ & & $-0.32$ & *** & $8.4\cdot 10^{-2}$ & *** \\
  $g_M$ & $7.5 \cdot 10^{-2}$ & * & $-6.3$ & *** & $1.6$ & *** \\
  $\gamma_N$ & $-2.1\cdot 10^{-3}$ & *** & $-4.2\cdot 10^{-2}$ & *** & $3.2\cdot 10^{-2}$ & *** \\
  $w_G$ & $-6.9$ & *** & $15.6$ & *** & $65.5$ & *** \\
  $\gamma_G$ & $-8.2\cdot 10^{-3}$ & *** & $-2.9\cdot 10^{-3}$ & * & $7.6\cdot 10^{-2}$ & *** \\
  $d_G$ &   $4.6\cdot 10^{-5}$ & *** & $5.0\cdot 10^{-4}$ & *** & $-5.2\cdot 10^{-4}$ & *** \\
  \hline
Observations & 5208  & & 5208 & & 5208 & \\ 
Adjusted R$^{2}$ & 0.40 & & 0.70 & & 0.41 & \\
\hline
\end{tabular}
\end{table}

Finally, as the output produced by the PSE algorithms are assumed to be mostly representative of what the model can offer, we can expect statistical models linking parameters and indicators to capture most of its behavior. We propose therefore in Table~\ref{tab:regpse} a linear regression analysis of model behavior. The estimation is done on the full PSE population but with weighting according to the number of stochastic samples, in order to avoid bias by non robust patterns. Most of variations explained in the grid experiment are confirmed, as hierarchies increasing with $d_G$, or decreasing with $\gamma_G$. The overall behavior of correlations is opposed to what was observed as a function of $d_G$ since it decreases. It is also interesting to note that centrality hierarchy and correlation are not significant in $\alpha_S$, while population hierarchy is not significant in $\phi_0^{(q)}$: on these dimensions, the intrication between cities and the transportation network is not statistically effective for a linear model (since these non-significant link occur between city indicator and network parameter on the one hand, and between network indicator and city parameter on the other hand).

\section{Discussion}



%

Our model exploration results have implications on the thematic question of hierarchy in urban systems and the role of co-evolution between cities and networks in its dynamics. We showed several stylized fact which have non trivial implications, including: (i) the fact that urban hierarchy depends on network processes, and that in some cases this link is non-monotonous - what introduces an additional complexity in planning such infrastructures at a macroscopic scale if put in a long time co-evolutionary context; (ii) the fact that correlation between urban hierarchy and network hierarchy are most of the time positive, but that it can take a broad range of values and even be negative - this also challenges the reductionist view of a direct correspondance between the hierarchy of a city and its accessibility, since the link depends of several parameters and of the type of interactions considered; (iii) the fact that conclusions obtained with the physical network model are globally qualitatively similar to the conclusions obtained with the virtual network, but that behavior still significantly differs in some regions of the parameter space for some indicators - what means that in some case such a simplification will be fine to be done while in other it will miss some crucial processes; (iv) the fact that the realm of possible hierarchy regimes is very broad, surely much broader than existing regimes.

This last point opens the issue of comparing this approach with data and possibly identifying hierarchy regimes in existing urban systems. \cite{raimbault2018modeling} have applied this model to real population data and real rail network distance matrices in the case of the French urban system, by calibrating it on population and distance trajectories. As the model is fitted on a moving window in time, the temporal trajectory of fitted parameters may inform on the actual regime the urban system is in. However, such conclusions would be more robust if applied on different urban systems, as \cite{raimbault2019evolutionary} do for six large urban systems when benchmarking similar interaction growth models. A purely empirical characterization of hierarchy regimes, using indicators introduced here, would also be a relevant entry to this issue, but the lack of transportation data on long time scales and broad spatial spans remains an obstacle difficult to overcome.




The methodology to understand hierarchical patterns in systems of cities, and the model itself are also prone to several potential developments. For example, the idea of spatial non-stationarity in estimating scaling laws, which would in a sense be linked to the existence of urban subsystems with their own hierarchical patterns, should be developed in methodological terms. An heuristic to optimize the adjustment of such a non-stationary model has to be introduced, and may be difficult to elaborate since a spatial neighborhood is not necessarily the rule in constituting subsystems of cities (large global metropolises may be a subsystem linked tighter than one of these with its hinterland in Europe for example). This then also relates to issues of relevant scales to identify hierarchies. Regarding the model in itself, it remains very simple and not realistic in the sense that similarly to \citep{xie2009topological}, no link are added, but only the speeds of existing links are updated. On the contrary, road network growth models at other scales such as \cite{raimbault2019urban} focus on the addition of links. Bridging these two approaches would be a relevant extension of the model studied here.




Finally, our results can be put into a wider theoretical perspective. As explained in introduction, hierarchies in the sense of the imbrication of subsystems at multiple levels, are endogenous to complex systems. At a fixed scale, quantitative indicators such as the one we used capture emergent patterns of this organisation, as is the hierarchical structure in systems of cities in terms of scaling laws. Thus, to understand and manage such systems in a resilient and adaptive way, multi-scale approaches embracing these hierarchies are necessary, as put forward out by \cite{rozenblat2018conclusion}. Our model is a first suggestion of scale integrations, since in the physical network case cities are at the macroscopic scale while the network is at a finer mesoscopic scales.

\section{Conclusion}

We explored here the concept of hierarchy in the particular context of the co-evolution of transportation networks and cities. More particularly, we introduced a set of indicators to quantify hierarchy patterns, and systematically studied a co-evolution model for cities and networks, at two abstraction levels for the network. Our exploration results provide some non trivial stylized facts and inform on the diversity of regimes the model can produce. This provides an illustration of how to study hierarchy in territorial systems regarding two complementary dimensions, in terms of how each hierarchically organizes and what is the actual correspondance between the two hierarchies.

\section{Acknowledgements}

Results obtained in this paper were computed on the vo.complex-system.eu virtual organization of the European Grid Infrastructure ( http://www.egi.eu ). We thank the European Grid Infrastructure and its supporting National Grid Initiatives (France-Grilles in particular) for providing the technical support and infrastructure. This work was funded by the Urban Dynamics Lab grant EPSRC EP/M023583/1.




\end{document}